\documentclass[12pt]{article}
\usepackage{jheppub}
\usepackage{enumerate}
\usepackage{epsfig}
\usepackage{amsmath} 
\usepackage{capt-of}
\usepackage{wasysym} 
\usepackage{mathrsfs} 
\usepackage{graphicx} 
\usepackage{amsfonts} 
\usepackage{amsbsy} 
\usepackage{amscd}
\usepackage{multirow}
\usepackage{rotating}
\usepackage{slashed, latexsym, verbatim, cancel} 
\usepackage[utf8]{inputenc}
\usepackage {multirow}
\usepackage[normalem]{ulem}
\usepackage{color}
\definecolor{myred}{rgb}{0.6,0,0} %usage:  {\textcolor{myred}{Hello World}}
\definecolor{myblue}{rgb}{0,0.2,0.4}
\definecolor{mygreen}{rgb}{0,0.9,0.1}
\definecolor{hc}{rgb}{.9,0.1,0.7}
\definecolor{hcout}{rgb}{.9,0.7,0.9}
\definecolor{Orange}{rgb}{1.,0.65,0.}
\makeatletter

\newcommand{\fmslash}[2][0mu]{%
  \mathchoice
    {\fmsl@sh\displaystyle{#1}{#2}}%
    {\fmsl@sh\textstyle{#1}{#2}}%
    {\fmsl@sh\scriptstyle{#1}{#2}}%
    {\fmsl@sh\scriptscriptstyle{#1}{#2}}}
\newcommand{\fmsl@sh}[3]{%
  \m@th\ooalign{$\hfil#1\mkern#2/\hfil$\crcr$#1#3$}}
  
  \newcommand\blfootnote[1]{%
  \begingroup
  \renewcommand\thefootnote{}\footnote{#1}%
  \addtocounter{footnote}{-1}%
  \endgroup
}

\makeatother

\newcommand{\lsim}{{\;\raise0.3ex\hbox{$<$\kern-0.75em\raise-1.1ex\hbox{$\sim$}}\;}}
\newcommand{\gsim}{{\;\raise0.3ex\hbox{$>$\kern-0.75em\raise-1.1ex\hbox{$\sim$}}\;}}

\usepackage{array}
\newcolumntype{C}[1]{>{\centering\arraybackslash$}p{#1}<{$}}

\usepackage{bm} % bold in math mode

\usepackage{cleveref}
\newcommand{\be}{\begin{equation}}
\newcommand{\ee}{\end{equation}}
\newcommand{\bes}{\begin{equation*}}
\newcommand{\ees}{\end{equation*}}
\newcommand{\bea}{\begin{eqnarray}}
\newcommand{\eea}{\end{eqnarray}}
\newcommand{\beas}{\begin{eqnarray*}}
\newcommand{\eeas}{\end{eqnarray*}}

\newcommand{\la}{\lambda}

\newcommand{\wt}{\widetilde}
\title{Searching for a Light (pseudo)Scalar via the Yukawa Process at the ILC}
\author{Eung Jin Chun,}
\author{Tanmoy Mondal$^\ast$\blfootnote{$^\ast$Corresponding author}\blfootnote{Talk presented at the International Workshop on Future Linear Colliders (LCWS2019), Sendai, Japan, 28 October-1 November, 2019. C19-10-28. This proceeding is based on ref~\cite{Chun:2019sjo}.}} 
\affiliation{Korea Institute for Advanced Study, Seoul 02455, Korea}
\emailAdd{ejchun@kias.re.kr}
\emailAdd{tanmoy@kias.re.kr}

\abstract{
Yukawa production of a light scalar can be explored at a linear collider. Light (pseudo)scalar can exist in extended Higgs models and an interesting example is the light pseudoscalar in Type-X two Higgs doublet model.  The model can explain the anomalous magnetic moment of muon at large $\tan\beta$. We show that the available parameter space in this model can be examined by the Yukawa process at $5\sigma$ at the ILC.
}
 \preprint{KIAS-P20008}
\date{\today}
\keywords{Two Higgs Doublet Models, Light (Pseudo)Scalar, Lepton Collider, ILC, Yukawa process, Higgs factory}

\begin{document}
\maketitle
\section{Introduction}
The two Higgs doublet model(2HDM) has been motivated by supersymmetry~\cite{Haber:1984rc}, baryon asymmetry of the Universe~\cite{Turok:1990zg,Trodden:1998ym} and the strong CP problem~\cite{Kim:1986ax}. To avoid the flavour changing neutral current(FCNC) processes four different types of 2HDM~\cite{Gunion:1989we,Djouadi:2005gj,Branco:2011iw} can be constructed with appropriate $\mathbb{Z}_2$ charge assignments. Among the different 2HDMs, the Lepton-specific or Type-X 2HDM is of particular interest as it can explain the observed anomaly of muon anomalous magnetic moment, ($g-2$)$_\mu$~\cite{Brown:2001mga,Bennett:2006fi}. The muon anomaly can be resolved in this model with a light pseudoscalar and large $\tan\beta$~\cite{Cao:2009as,Broggio:2014mna,Ilisie:2015tra,Abe:2015oca,Chun:2016hzs,Cherchiglia:2017uwv,Wang:2018hnw}.

Since the light pseudoscalar($A$) in Type-X 2HDM model is leptophilic, it is very hard to produce $A$ via gluon fusion at the large hadron collider (LHC). Hence, the model can be explored at the LHC via the associated production of $A$ along with a charged ($H^\pm$) or neutral ($H$) scalar~\cite{Chun:2015hsa, Chun:2018vsn} and from the decay of the 125 GeV SM Higgs boson ($h$) into a pair of pseudoscalars \cite{Chun:2017yob}. However, these searches depend on the additional parameters like the Higgs to AA branching or the masses of the heavy scalars.

A light pseudoscalar can be searched at a lepton collider via the Yukawa process where a light $A$ is radiated from a tau lepton. Any new lepton collider will run as \emph{Higgs factory} with the center-of-mass energy ($\sqrt{s}$) close to 250 GeV. Hence it is important to study the prospect of the search for a light boson at a 250 GeV lepton collider like ILC~\cite{Baer:2013cma, Bambade:2019fyw} which has not been studied before.  The Type-X 2HDM 
 model was studied via $4\tau$  and $2\mu2\tau$ channel at 500 GeV and 1 TeV lepton colliders where the associated production is the dominant mode~\cite{Kanemura:2012az,Hashemi:2017awj}.

Here we are interested in the Yukawa production of a light pseudoscalar motivated by the $(g-2)_\mu$ measurement. 
Another novel feature of the Yukawa production is that the process does not depend on the masses of the heavy scalars present in the theory. In this article we explore how to search for such a particle at a \emph{Higgs factory} in the Yukawa channel with four tau final state. We apply the collinear approximation to reconstruct the mass of the light pseudoscalar unambiguously. We found that 
ILC with 2000 $fb^{-1}$ of integrated luminosity can explore the whole $(g-2)_\mu$ compatible parameter space of the  
Type-X 2HDM.

\section{The Type-X 2HDM Model}\label{sec:model}
The 2HDM model has been discussed in detail in ref~\cite{Gunion:1989we,Djouadi:2005gj,Branco:2011iw}. 
The model consists of two scalar doublets($\Phi_1 \& \,\Phi_2$) with hypercharge +1. In general, both the doublets can
couple to the fermions which leads to FCNC interaction at tree level. To avoid this, we impose an
additional $\mathbb{Z}_2$ symmetry  such that $\Phi_1\rightarrow-\Phi_1$ and $\Phi_2\rightarrow \Phi_2$. The scalar potential then reads as,
\begin{eqnarray}
\nonumber V_{\mathrm{2HDM}} &=& -m_{11}^2\Phi_1^{\dagger}\Phi_1 - m_{22}^2\Phi_2^{\dagger}\Phi_2 -\Big[m_{12}^2\Phi_1^{\dagger}\Phi_2 + \mathrm{h.c.}\Big]
+\frac{1}{2}\lambda_1\left(\Phi_1^\dagger\Phi_1\right)^2+\frac{1}{2}\lambda_2\left(\Phi_2^\dagger\Phi_2\right)^2 \\
\nonumber && +\lambda_3\left(\Phi_1^\dagger\Phi_1\right)\left(\Phi_2^\dagger\Phi_2\right)+\lambda_4\left(\Phi_1^\dagger\Phi_2\right)\left(\Phi_2^\dagger\Phi_1\right)
+\Big\{ \frac{1}{2}\lambda_5\left(\Phi_1^\dagger\Phi_2\right)^2+  \rm{h.c.}\Big\}.
\label{eq:2hdm-pot}
\end{eqnarray}
The mass term $m_{12}^2$ softly breaks the $\mathbb{Z}_2$ symmetry and we have assumed that all the 
couplings are real. 
After the electroweak symmetry breaking we can parameterize the doublets in the following way,
$\Phi_j=(H_j^+,(v_j + h_j + i A_j)/\sqrt{2})^T$ where $v_j$ denotes vacuum expectation values. We can write the massive physical states $A$ (CP-odd), $h$, $H$, $H^{\pm}$ in terms of the gauge eigenstates:
\begin{align*}
 \begin{pmatrix} H  \\ h \end{pmatrix} =  \begin{pmatrix}  c_{\alpha} && s_{\alpha} \\ -s_{\alpha} &&  c_{\alpha} \end{pmatrix}
  \begin{pmatrix} h_1  \\ h_2 \end{pmatrix},
  \quad
  A=-s_\beta \;A_1 + c_\beta \;A_2\,,  \textrm{ and }\;\; H^{\pm}=-s_\beta\; H_1^{\pm} + c_\beta\; H^{\pm}_2,
 \end{align*}
where $s_\alpha = {\rm sin}~\alpha$, $c_\beta = {\rm cos}~ \beta$ etc and ${\rm tan}~\beta = \cfrac{v_2}{v_1}$ .
The lightest CP-even eigenstate $h$ is identified with the SM-like Higgs with mass $m_h \approx 125$ GeV.

There are four possible type of Yukawa structures in 2HDM models depending on the $\mathbb{Z}_2$ charge assignment of the fermions. In  this article we will consider the Type-X 2HDM where the RH leptons are odd under the $\mathbb{Z}_2$ symmetry. The relevant Yukawa Lagrangian is given by,
\begin{equation}\label{eq:yukawa}
-{\cal L}_Y= Y^u\bar{ Q_L} \wt \Phi_2 u_R + Y^d  \bar{ Q_L} \Phi_2 d_R+Y^e\bar{ l_L} \Phi_1 e_R + h.c.,
\end{equation}
where $\wt \Phi_2=i\sigma_2\Phi_2^*$. After symmetry breaking the we can write the Yukawa Lagrangian in terms of mass eigenstates,
\begin{eqnarray}
\nonumber \mathcal L_{\mathrm{Yukawa}}^{\mathrm{Physical}} &=&
-\sum_{f=u,d,\ell} \frac{m_f}{v}\left(\xi_h^f\overline{f}hf +
\xi_H^f\overline{f}Hf - i\xi_A^f\overline{f}\gamma_5Af \right) \\
 &&-\left\{ \frac{\sqrt{2}V_{ud}}
{v}\overline{u}\left(\xi_A^{u} m_{u} P_L+\xi_A^{d} m_{d} P_R\right)H^{+}d  +
\frac{\sqrt{2}m_l}{v}\xi_A^l\overline{v}_LH^{+}l_R + \mathrm{h.c.}\right\},
\label{eq:L2hdm}
\end{eqnarray}
where  $u$, $d$, and $l$ refer to the up-type quarks, down-type quarks, and charged leptons, respectively. 
The multiplicative factors, {\it i.e.} $\xi_{\phi}^f$ are given in Table~\ref{Tab:YukawaFactors}. 
When $\cos(\beta-\alpha) \to 0 $, the modifiers to the SM like Higgs($\xi_{h}^\ell$) becomes +1 and matches with the SM Yukawa coupling. This limit 
is called right sign (RS) Yukawa limit. On the other hand, the modifier $\xi_{h}^\ell$  goes to `-1' 
 if $\cos(\beta-\alpha)$ takes the value $2/\tan\beta$ and consequently, this limit of $(\beta-\alpha)$ is known as wrong sign (WS) Yukawa limit.
%   
%  The decay width of the pseudoscalar $A$ to a pair of fermions is given by,
%  \be\label{eq:decaywidth}
%  \Gamma(A\to f\bar{f}) = \frac{1}{8\pi} \left(\frac{m_f }{v} \xi_A^f \right)^2 m_A \sqrt{1-\left(\frac{2 m_f}{m_A}\right)^2}.
%  \ee
% Since $\xi_A^f =\tan\beta$, $A$ will decay dominantly into tau leptons with a small branching fraction to muons ($BR(A\to\mu\mu)\simeq0.0035$) when $\tan\beta$ is large. Since we are interested in large $\tan\beta$ region, we will 
% consider only $A\to\tau\tau$ final states.
\begin{table}[t]
%-------------------------------------------------------------------------------
\begin{center}
\begin{tabular}{|c||c|c|c|c|c|c|c|c|c|}
\hline
& $\xi_h^u$ & $\xi_h^d$ & $\xi_h^\ell$
& $\xi_H^u$ & $\xi_H^d$ & $\xi_H^\ell$
& $\xi_A^u$ & $\xi_A^d$ & $\xi_A^\ell$ \\ \hline
Type-X
& $c_\alpha/s_\beta$ & $c_\alpha/s_\beta$ & $-s_\alpha/c_\beta$
& $s_\alpha/s_\beta$ & $s_\alpha/s_\beta$ & $c_\alpha/c_\beta$
& $\cot\beta$ & $-\cot\beta$ & $\tan\beta$ \\
 \hline
\end{tabular}
\end{center}
 \caption{The multiplicative factors of Yukawa interactions in type X 2HDM}
\label{Tab:YukawaFactors}
\end{table}

\subsection{Constraints on the model}

 Vacuum stability, perturbativity and unitarity put constraints on the quartic couplings. The following constraints should be 
 satisfied~\cite{Broggio:2014mna,Wang:2014sda}:
\bea
m_{H} \simeq m_{H^\pm} &\leq& 250 \textrm{ GeV} \hspace{1cm} \textrm{(RS scenario)}\nonumber \\
m_{H} \simeq m_{H^\pm} &\leq& \sqrt{\la_{max}}\ v = \sqrt{4\pi}\ v\hspace{1cm} \textrm{(WS scenario)}.
\eea 
We will appropriately choose the value of $\cos(\beta-\alpha)$ to satisfy these conditions. The constraints from the electroweak precision measurements require that the charged Higgs boson has to be nearly degenerate with either $H$ or $A$~\cite{Broggio:2014mna, Haller:2018nnx}. We will assume $m_{H}\simeq m_{H^\pm}$ to satisfy the 
EWPT. 

Heavy scalar searches at the LHC has very little impact on 2HDM-X parameter space as the scalars in this model are hadrophobic and their coupling to the quarks decreases as $\tan\beta$ increases. The limit from LEP on pair production of $A$ and $H$ via $Z$ is $m_A+m_H  > 185 $ GeV~\cite{Abdallah:2004wy} which we will respect. Since the new scalars couple to quarks very weakly the flavour constraints coming from $B\to X_s\gamma$ or $ B_s\to\mu^+\mu^-$ are weak and for $\tan\beta > 5$, there is no limit on the scalar spectrum~\cite{Haller:2018nnx}. 
The global analysis of the present Higgs data allows the WS limit for large $\tan\beta$~\cite{Haller:2018nnx}. Due to the leptophilic nature of such a light pseudoscalar, lepton universality tests \cite{ALEPH:2005ab, Amhis:2016xyh} 
can provide severe bounds on the parameter space
favorable for $(g-2)_\mu$ \cite{Abe:2015oca,Chun:2016hzs}.

The parameter space we are interested is where the pseudoscalar is light (i.e. $m_A < 90$ GeV)  and $\tan\beta$ is large. 
For simplicity, we assume $m_H=m_{H^\pm}=250$ GeV. 
We also choose the wrong-sign Yukawa limit because the $BR(h\to AA)$ can be small~\cite{Chun:2015hsa} in this limit 
which can satisfy the present LHC bound~\cite{Sirunyan:2018mbx}.

\section{Search for Yukawa process at lepton collider}\label{sec:simulation}
\begin{figure}[!t]
\begin{center}
 \includegraphics[width=7cm,angle=270]{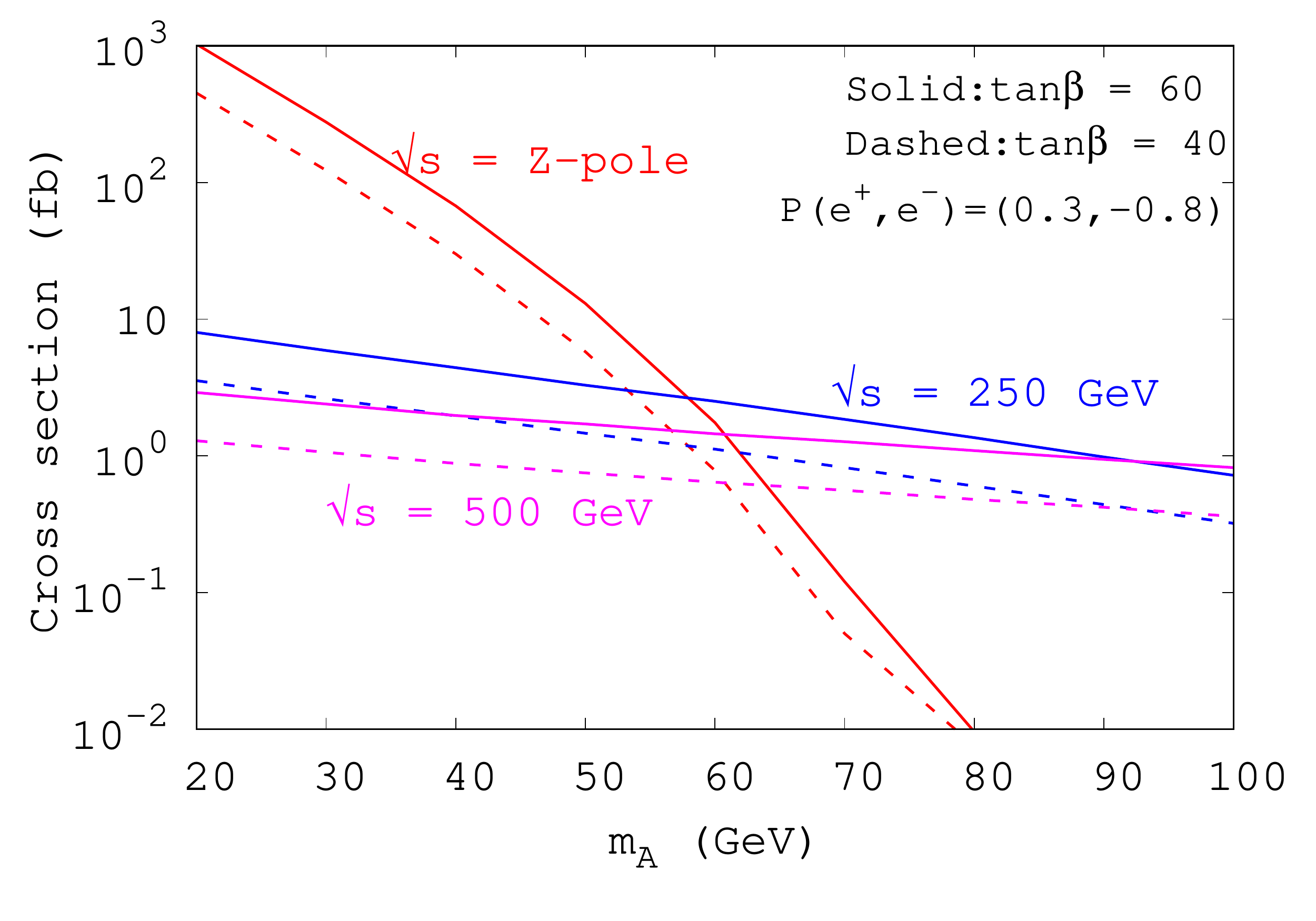}
 \caption{Production cross section of $e^+ e^- \to Z^*\gamma^* \to \tau\tau A$ as a function of the light boson mass at different center-of-mass energy. }
 \label{fig:x-section}
\end{center}
\end{figure}

The Yukawa process under the consideration is, 
$$e^+ e^- \to Z^*/\gamma^* \to \tau^+ \tau^- A \to  4 \tau.$$ 
The production cross section of $4\tau$ as a function of $m_A$ for different $\sqrt{s}$ is depicted in 
Fig.~\ref{fig:x-section} where we have used polarized beam with $P(e^+,e^-)=(+30\%,-80\%)$~\cite{Behnke:2013lya}. 
Since $A\tau\tau$ coupling is proportional to $\tan\beta$, cross-section increases as $\tan\beta$ increases. 
Although it is easier to produce a light $A$ at $Z$-pole, the taus originating from the decay of $A$ will be soft and will remain undetected. The  250 GeV centre-of-mass energy is perfect to explore the Yukawa structure.
The signal events are identified  as 
$$3 \ j_\tau + X, \hspace{1cm} X \equiv j_\tau\ /j \ /\ell_\tau,$$ where $j_\tau$ is a $\tau$-tagged jet; $j$ is an  untagged jet, 
and $\ell_\tau \equiv e/\mu$ is leptons from the decay of $\tau$ such that the total number of objects is four. The inclusion of a lepton 
in the final state helps to increase signal events since leptonic decay of a tau is substantial.
 
 The background to this channel comes from the $e^+e^- \to Z(\gamma^*) \; Z(\gamma^*) \to 4\tau$ and 
 $e^+e^- \to Z(\gamma^*) \; Z(\gamma^*) \to 2\tau \ 2 j$ 
 processes where mis-identification of a light jet into a $\tau$-tagged jet mimics the signal in the latter case. There are subdominant background 
 coming from the $e^+e^- \to Z h$ process. At 250 GeV ILC we estimated that the total parton level production cross-section of 
 $4~\tau$ background process is $\sim 6.6 fb$  and the cross-section for the $2\tau\ 2j$ process is $\sim 250 fb$. 

\subsection{Event simulation and selection}

 We have used $\texttt{MadGraph5\_aMC@NLO}$ \cite{Alwall:2011uj,Alwall:2014hca} to produce the parton level signal and background events and then used \texttt{PYTHIA8} \cite{Sjostrand:2006za,Sjostrand:2014zea} for the subsequent decay, showering and hadronization. The $\tau$ decays 
are incorporated in $\texttt{MadGraph5\_aMC@NLO}$ via \texttt{TAUOLA}~\cite{Jadach:1993hs}. To simulate the detector effects we have used \texttt{Delphes3} \cite{deFavereau:2013fsa} with the ILD detector card. Jets are clustered using the longitudinal-kT algorithm~\cite{Catani:1993hr,Ellis:1993tq} with $R = 0.4$. At the LHC analysis the tau tagging efficiency of 
$\tau$ jets($\epsilon_\tau$) is 60\% is used in general~\cite{CMS-PAS-TAU-16-002}. However, since ILC is a lepton collider it is expected to have better tagging efficiency for jets due to improved track momentum and jet energy resolution. Hence we have considered two different tau tagging efficacy($\epsilon_\tau$): one conservative $\epsilon_\tau = 60\%$  and one optimistic $\epsilon_\tau = 90\%$~\cite{Jeans:2018anq}. We set mis-tagging rate at 0.5\% for both the cases. We have used the Delphes jet charge measurement to make opposite sign jet pair.
 
We imposed the \textit{pre-selection criteria} that all the jets and leptons should have minimum energy of 20 GeV and should have $|\eta| <2.3$ which corresponds to $|\cos\theta| < 0.98$. Using the selected events we then move on to reconstruct the parent $\tau$-leptons.

\subsection{Collinear approximation and reconstruction of A}
The collinear approximation assumes that the missing energy from the decay of tau lepton is collinear to the visible part of the decay.
This approximation is true when tau lepton is boosted enough and in the Yukawa process discussed here, the energy spectrum is in general hard. Using this approximation it is possible to reconstruct the momentum of the four taus. The energy momentum conservation equations are,
\beas
\vec{p}(\tau_1)+\vec{p}(\tau_2)+\vec{p}(\tau_3)+\vec{p}(\tau_4) &=& \vec{0},\\
E(\tau_1)+E(\tau_2)+E(\tau_3)+E(\tau_4) &=& \sqrt{s}.
\eeas
 
 Let us assume that the $i$-th object from $\tau$ decay takes $z_i$ amount of the original momenta $i.e.$ $p^\mu(j_i) = z_i  \ p^\mu (\tau_i)$. Given four visible four momentum we can solve the above set of equations for $z_i$. The physical solutions ensures  
 that $0 < z_i < 1$. However, due to finite momentum resolution of the jets and since we are dealing with at least 3 $\tau$-tagged jets in the final state there will be some uncertainty in the solution. Accordingly, we  have relaxed the condition on $z_i$ such that  $z_i < 1.1$~\cite{Kanemura:2011kx}. Now using the $z_i$ we can reconstruct the momentum of the tau-leptons and finally reconstruct the pseudoscalar. 
 
  To identify the $A$ resonance without any ambiguity we have to find out the opposite sign tau-pair originating from the pseudoscalar.  Since there are four $\tau$s, there will be four possible opposite sign tau-pair and we use the following method to find out the correct combination:

\begin{figure}[t!]
\begin{center}
 \includegraphics[width=6.8cm]{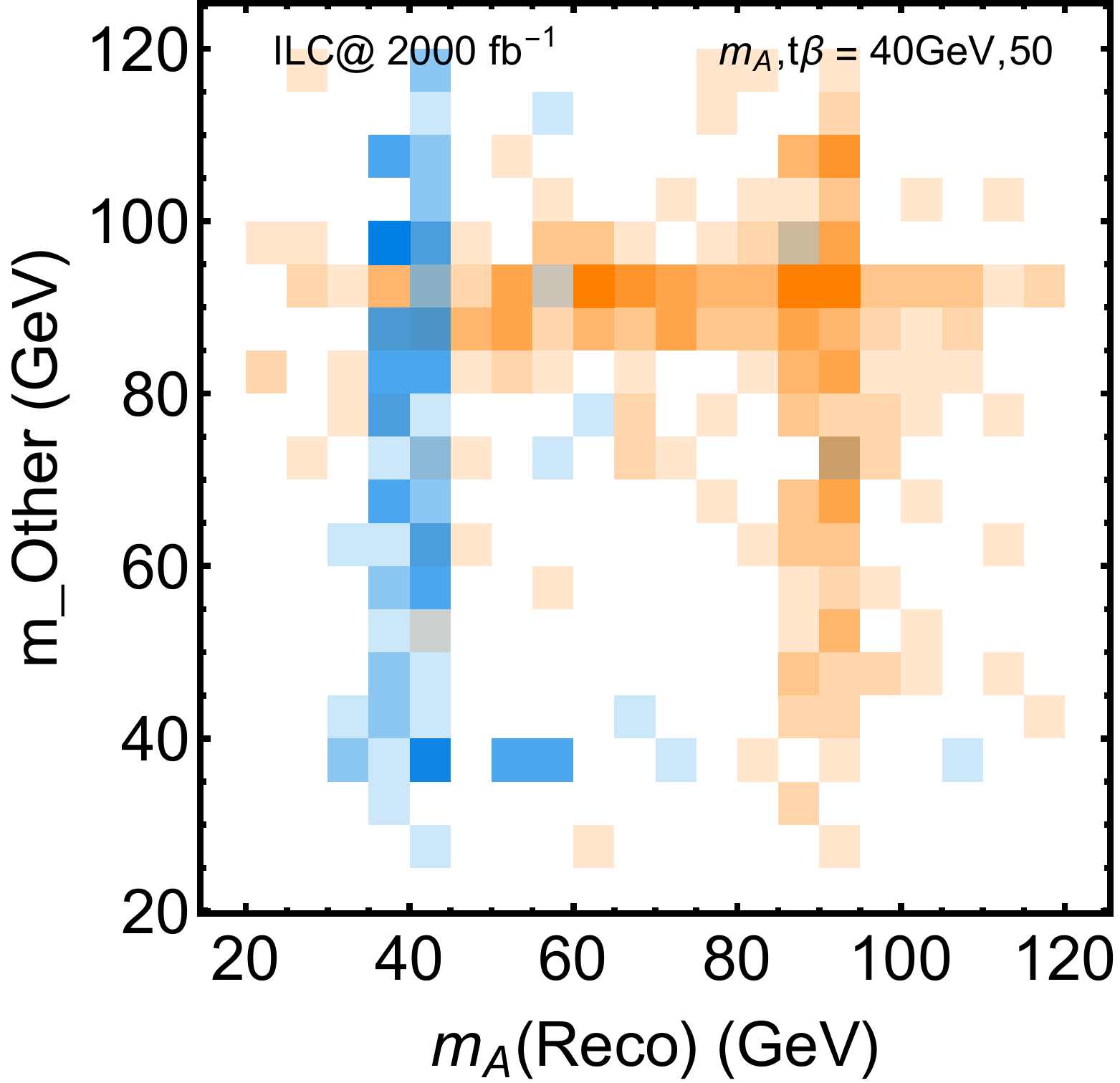}\hskip15pt
 \includegraphics[width=8.2cm]{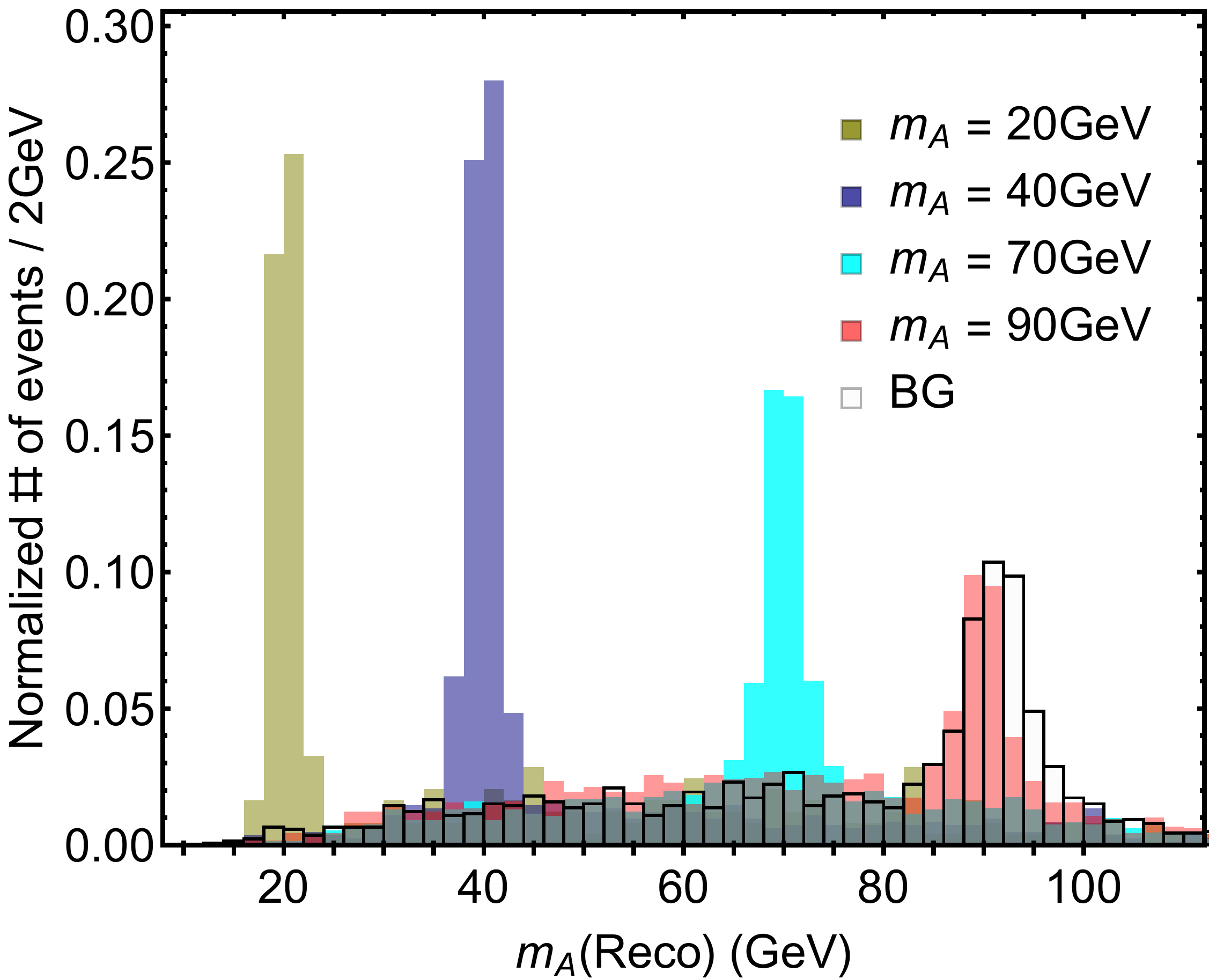}
\end{center}
\caption{\emph{Left Panel:} Density plot of $m_A(Reco)$ and $m_{\rm{Other}}$ for signal benchmark point ($m_A = 40$ GeV and $\tan\beta =50$) in blue and for background events in orange. See text for definitions of invariant masses. Signal and background events are 
generated at $\sqrt{s}=$250 GeV with ILC environment with integrated luminosity of 2000 $fb^{-1}$. \emph{Right panel:} Normalized invariant mass distribution of the reconstructed pseudoscalar using collinear approximation for different pseudoscalar mass.}
\label{fig:inv_mass_all}
\end{figure}

 \begin{itemize}
 \item From the three-body decay kinematics we know that the maximum available energy for $A$ is varies from 126 GeV(for $m_A =20$ GeV) to 141 GeV(for $m_A=90$GeV). On the other hand, the energy of the $\tau$s produced in association with the $A$ can reach close to $\sqrt{s}/2$. Hence is it reasonable to assume that the highest energy tau is coming from the decay of $Z$, not from the radiated $A$. We remove the highest energy tau.
 \item From the remaining three taus we can construct two possible opposite sign combinations. 
 \item Between these two combinations,  the $\tau$-pair which gives the highest transverse momentum($p_T$) is likely to come from the decay of $A$. We calculate the invariant mass from this combination which is denoted as $m_A(Reco)$.
 The invariant mass from the other opposite sign tau pair is denoted as $m\_{Other}$. 
\end{itemize}

To show the effectiveness of the method, we have plotted $m_A(Reco) \ \& \ m\_{Other}$ in the left panel of Fig.~\ref{fig:inv_mass_all} for  pseudoscalar mass of 40 GeV with $\tan\beta=50$. The signal events are displayed in blue, and the background events are shown in orange. The events are generated at 250 GeV ILC with integrated luminosity 
amount to 2000 $fb^{-1}$. The $m_A(Reco)$ clustered around true $A$ mass, i.e. near 40 GeV, whereas $m\_{Other}$ is scattered. Expectedly, the background events are clustered near the $Z$-mass, as the dominant background is from $ZZ$.  In the right panel of Fig.~\ref{fig:inv_mass_all} we present the reconstructed invariant mass distribution $m_A(Reco)$ for different values of $m_A$. As $m_A$ increases, the invariant mass peak becomes broader since the decay width of $A$ is proportional to its mass.

\section{Results}\label{sec:result}
\begin{table}[t!]
\begin{center}
\begin{tabular}{|c||c|c|c||c|}
\hline
 \multicolumn{5}{|c|}{\bf Pre-selection cut : Energy $>$ 20 GeV. $|\eta| <$ 2.3 }\\
\hline
$\mathcal{L}$ = 2000  $fb^{-1}$ & \multirow{2}{*}{Signal} & \multicolumn{2}{c|}{Background}   & \multirow{2}{*}{Significance}   \\ \cline{1-1} \cline{3-4} 
                               &                         & 4$\tau$ & 2$\tau$ \ 2 j &                \\ \hline
Pre-selection cut               & 106 [100\%]            & 242 [100\%]       & 98[100\%]            & 5.5 \\ \hline
Collinear approx  &&&&\\
$0 < z_i <1.1$               & 91 [86.0\%]           & 217[89.7\%]      & 69[70.4\%]         & 5.1 \\ \hline
$m_A\pm10$GeV                  & 66 [62.3\%]           & 32   [14.9\%]       & 10[10.2\%]            & 8.5  \\ \hline
\end{tabular}
\caption{Cut flow for $m_A = 40$~GeV and $\tan \beta=50$ with integrated luminosity of 2000 $fb^{-1}$. } 
\label{tab:cut-fow}
\end{center}
\end{table}

\begin{figure}
\begin{center}
    \includegraphics[width=9cm]{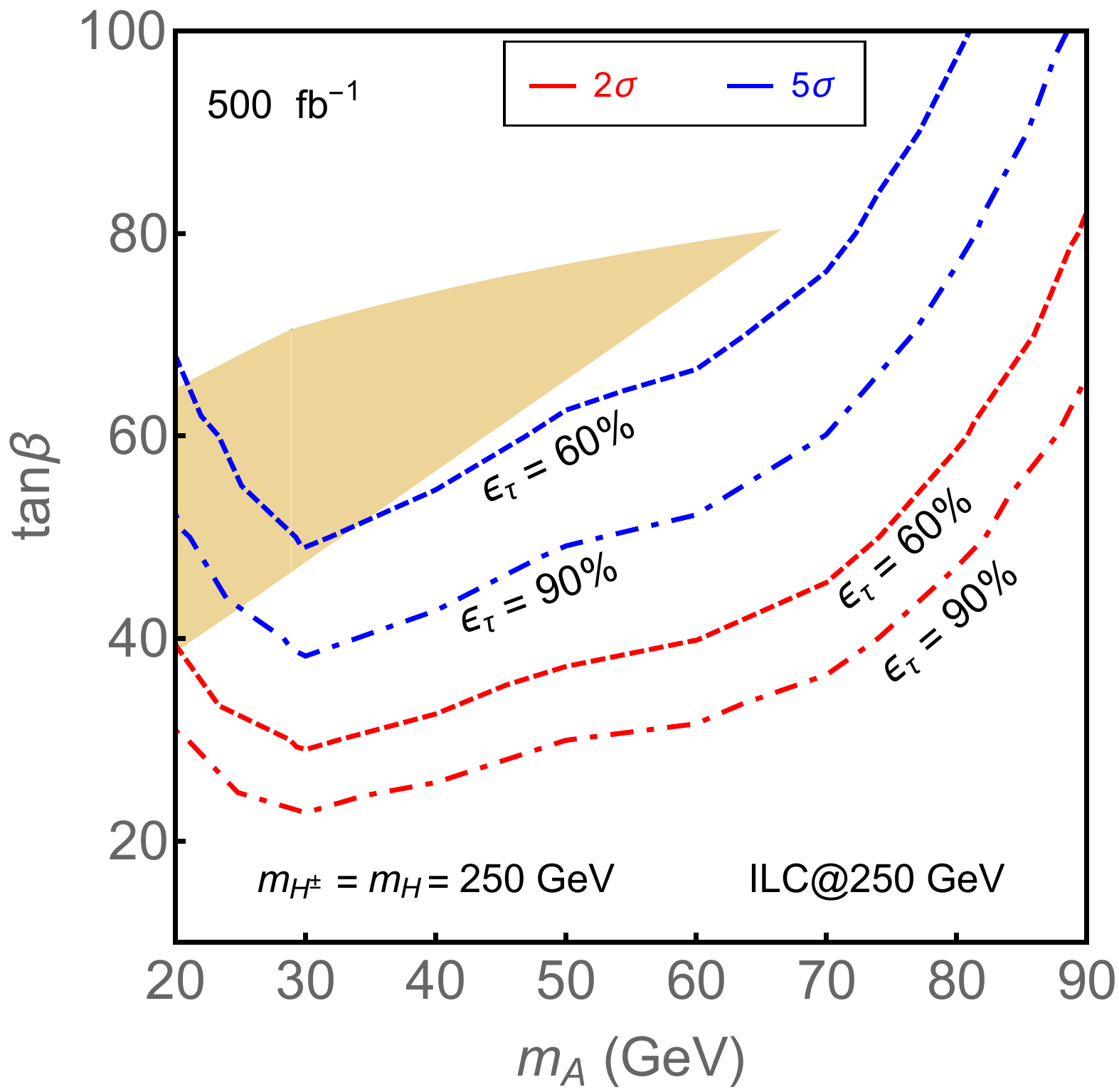}
\end{center}
\caption{Reach of the 250 GeV $e^+e^-$ collider in the $m_A$--$\tan\beta$ parameter space of the Type-X 2HDM with 
$\epsilon_\tau$=60\% and 90\%. The significance increases for larger $\tan\beta$ as the 
signal production cross section is almost proportional to $\tan^2\beta$. The light yellow region 
can explain the $(g-2)_\mu$ anomaly at $2~\sigma$ after applying the lepton universality constraints.
}
\label{fig:param-scan}
\end{figure}

In the previous section, we have established that the collinear approximation can be used to reconstruct the mass of $A$. 
Now, we will use the reconstructed invariant mass to minimize the background events and enhance the signal efficiency.
 The cut flow table for one benchmark ($m_A = 40$ GeV and $\tan\beta =50$) and background events is shown in 
Table.~\ref{tab:cut-fow} with integrated luminosity of 2000 $fb^{-1}$. The background cross-section is low, and it is possible to achieve large-signal significance at the pre-election level and after using the invariant mass cut the significance increased to more than  $8~\sigma$. The significance was calculated using the well-known expression
\begin{equation}
\mathcal{S} = \sqrt{2\left[(S+B)\textrm{ln}\left(1+\frac{S}{B}\right)-S\right]}, 
\end{equation}
 where $S(B)$ denotes number of signal (background) events after the cuts. 

We have scanned full the $m_A - \tan\beta$ parameter space and compute the signal significance at ILC250 with integrated luminosity of 500 $fb^{-1}$ and 2000~$fb^{-1}$. 
In Fig.~\ref{fig:param-scan} we have shown $2~\sigma$ exclusion and $5~\sigma$ discovery contours.
Here, the dashed(dot-dashed) lines are contours with $\epsilon_\tau = 60\%(90\%)$ and the red(blue) lines are contours with 2$\sigma(5\sigma)$ significance. As $m_A$ decreases, the decay products become soft, which leads to weak bound, and at higher $m_A$ the signal cross-section decreases and background events increases which leads to weaker bounds. The allowed parameter space which can explain $(g-2)_\mu$ after satisfying the lepton universality constraints is shown in yellow. This result is obtained by following the analysis in \cite{Chun:2016hzs} and the updated in \cite{Chun:2019oix}. A large portion of the parameter space favoured by the muon anomaly can be explored at ILC250 even with 500 $fb^{-1}$ luminosity. With higher luminosity, the whole parameter space can be explored for this model.

\section{Conclusion}\label{sec:conclusion}

A light pseudoscalar in Type-X 2HDM at large $\tan\beta$ can explain the observed deviation of the muon anomalous magnetic moment, and it is worthwhile to test the scenario at hadron or lepton colliders like LHC or ILC. It is very hard to look for a leptophilic light pseudoscalar at LHC unless the heavier Higgs bosons, $H^\pm$ and $H$, are lighter than 200 GeV. On the other hand, lepton colliders are ideal to probe the parameter space via the (tau) Yukawa process independent of the heavy Higgs masses.

We demonstrated that in a \emph{Higgs-factory}, i.e. ILC 250, it is possible to test the model independent of the heavier scalar particles. We have done a realistic analysis with $4 \tau$ final states to reconstruct the light pseudoscalar by using the collinear approximation. The entire relevant parameter space compatible with the muon $(g-2)$ anomaly can be explored at $5~\sigma$ with an integrated luminosity of 2000 $fb^{-1}$.

\section*{Acknowledegments}
TM is supported by a KIAS Individual Grant (PG073501) and EJC is supported by a KIAS Individual Grant (PG012504) 
at Korea Institute for Advanced Study.

\bibliographystyle{JHEP}
\bibliography{2hdm}

\end{document}